\documentclass{jfm}
\usepackage[utf8x]{inputenc}
\usepackage[dvips]{color}
\usepackage{graphicx}
\usepackage{natbib}
\usepackage{url}

\newcommand{\beq}{\begin{equation}}  
\newcommand{\eeq}{\end{equation}}  
\newcommand{\bea}{\begin{eqnarray}}  
\newcommand{\eea}{\end{eqnarray}}

\title[Symmetry related dynamics in parallel shear flows]%
{Symmetry related dynamics\\ 
in parallel shear flows}
\author[T. Kreilos, S. Zammert and B. Eckhardt]%
{Tobias Kreilos$^{a,b}$\footnote{tobias.kreilos@physik.uni-marburg.de},
Stefan Zammert$^{a}$\footnote{stefan.zammert@physik.uni-marburg.de},
and Bruno Eckhardt$^{a,c}$\footnote{bruno.eckhardt@physik.uni-marburg.de}}
\affiliation{
$^a$Fachbereich Physik, Philipps-Universit\"at Marburg, Renthof 6, D-35032 Marburg, Germany\\ 
$^b$ Max Planck Institute for Dynamics and Self-Organization, Am Fassberg 17, D-37077 G\"ottingen, Germany\\
$^c$J.M. Burgerscentrum, Delft University of Technology, Mekelweg 2, 2628 CD Delft, The Netherlands}

\date{\today}

\begin{document}

\maketitle

\begin{abstract}
Parallel shear flows come with continuous symmetries of translation in the downstream
and spanwise direction. As a consequence, flow states that differ in their spanwise or downstream
location but are otherwise identical are dynamically equivalent. In the case of 
travelling waves, this trivial degree of freedom can be removed by going to 
a frame of reference that moves with the state, thereby turning the travelling
wave in the laboratory frame to a fixed point in the comoving frame of reference.
We here discuss a general method by which the 
translational displacements can be removed also for more complicated
and dynamically active states and  demonstrate its application for several examples.
For flows states in the asymptotic suction boundary layer we show that in the case of
the long-period oscillatory edge state we can find local phase
speeds which remove the fast oscillations and reveal the slow
vortex dynamics underlying the burst phenomenon. For spanwise translating states
we show that the method removes the drift but not the dynamical events that cause the
big spanwise displacement. For a turbulent case we apply the method to the 
spanwise shifts and find slow components that are correlated over very long times.
Calculations for plane Poiseuille flow show that the long correlations in the transverse motions are not 
special to the asymptotic suction boundary layer.
\end{abstract}


\section{Introduction}
  \label{sec:introduction}
Parallel shear flows like plane Couette flow or the asymptotic suction boundary
layer, have continuous translational symmetries in the wall-parallel directions:
flow states that differ only by a shift in these directions are dynamically equivalent.
Different geometries can bring in other or additional symmetries. For instance, convection 
in cylindrical containers is invariant under rotations around the cylinder axis, and flow down
a cylindrical pipe is invariant under rotations around the axis as well. The time evolution
of these flow fields does not always preserve these symmetries, in the sense that deformations
of the velocity fields and translations along the symmetric directions are mixed. The implications
of these continuous symmetries have been discussed in the context of 
exact coherent structures and dynamical systems approaches 
to the transition to turbulence in shear flows \citep{Willis2013,Cvitanovic2012,Mellibovsky2012}, 
but they are also related to the applications and analyses
using Taylors frozen flow approximations \citep{Taylor1938,Townsend1980,Zaman1981,DelAlamo2009}. 
In both cases the issue is whether and how one can remove the motion along the symmetry axis.

As a specific example for the connection to exact coherent structures, consider the case of plane 
Couette flow in small domains that are periodic in the downstream and spanwise direction
\citep{Nagata1990,Clever1997,Waleffe2003,Wang2007,Schneider2008}.
The exact coherent states described by \citet{Nagata1990,Clever1997}
are stationary solutions to the incompressible  Navier-Stokes equation. They are fully 3-d with all
velocity components active and variations in all three spatial directions, and they are invariant under translation
in the downstream and spanwise direction (because of the periodic boundary conditions). 
Accordingly, they can be shifted anywhere within the domain. 
A second class of states are travelling waves, i.e. states
that move downstream without change of shape, so that they can be described by 
velocity fields of the form ${\bf u}_{TW}(x-ct,y,z)$ \citep{Nagata1997,Gibson2008}. 
They become stationary in a frame of
reference moving with the phase speed $c$. The third class of states are periodic solutions
where the velocity field returns to its original shape after some period $T$.
More generally, the velocity field can reappear in shape but at a location displaced
along the symmetry axis \citep{Viswanath2007,Gibson2009}.

In the following we will outline calculations by which the advection
along a symmetry axis can be  extracted and removed from the time evolution, so 
that the non-trivial parts of the 
dynamics are highlighted. We will demonstrate the method for flow states in the 
asymptotic suction boundary layer \citep{Schlichting1987}, 
where rich dynamical structures exist \citep{Kreilos2013,Khapko2013,Khapko2014}, and 
in plane Poiseuille flow. Further examples and applications are forthcoming.

In \S2 we describe the method and its relation to other expressions in the literature. 
In \S3 we describe its application to states in the 
asymptotic suction boundary layer: the long-period edge state with its conspicuous bursts
\citep{Kreilos2013}, the sideways traveling localized states \citep{Khapko2013} and a turbulent
state, where slow spanwise drifts that are correlated over very long times are identified. 
In \S4 briefly discuss an application to plane Poiseuille flow, and show that the spanwise
drifts are also present in that case.  We conclude
with general remarks and an outlook in \S5.

\section{Symmetry related motions}
We consider the time evolution of a velocity field in the form
\beq
\partial_t{\bf u}(x,y,z,t) = {\bf f}({\bf u}(x,y,z,t), t)
\eeq
in a situation where the right hand side is invariant under a translation
along the $x$-axis.
The velocity field ${\bf u}(x,y,z,t)$ at time $t$ is one point in the state space
of the system, the velocity fields ${\bf u}(x,y,z,t+\Delta t)$ at time $t+\Delta t$
and ${\bf u}(x-\Delta x,y,z,t)$ displaced by $\Delta x$ in space are two
other points. The difference vectors in the direction of time evolution and spatial shifts
will be referred to as the time-evolution and displacement vectors, respectively. 
Using a first order Taylor-expansion, one finds that the time-evolution
vector points along the right hand side of the evolution equation,
\beq
{\bf u}(x,y,z,t+\Delta t) - {\bf u}(x,y,z,t) \approx  \Delta t\, {\bf f}({\bf u}(x,y,z,t), t)
\eeq
and the displacement vector points along the derivative of the velocity field,
\beq
{\bf u}(x-\Delta x,y,z,t) - {\bf u}(x,y,z,t) \approx -  \Delta x\, \partial_x {\bf u}(x,y,z,t)
\eeq
Usually, ${\bf f}$ has components parallel to the displacement vector, so that the
time evolution mixes a change in the velocity field with a translation. Removing the
displacement from ${\bf f}$ then leaves the non-trivial components of the time-evolution.
In order to relate $\Delta x$ and $\Delta t$, we introduce a phase velocity $c$ such that
$\Delta x = c\Delta t$, and then consider the time evolved and displaced field
${\bf u}(x-c \Delta t,y,z,t+\Delta t)$, which in a Taylor expansion to first order in time, becomes
\beq
{\bf u}(x- c \Delta t,y,z,t+\Delta t) \approx {\bf u}(x,y,z,t)+(- c  \partial_x {\bf u}(x,y,z,t) + 
{\bf f}({\bf u}(x,y,z,t), t))\Delta t
\eeq
Projecting with  $\partial_x {\bf u}(x,y,z,t)$ one finds that the parallel components are
removed by the choice
\beq
\label{eq:c}
c=\frac{ \langle \partial_x {\bf u}(x,y,z,t)\cdot {\bf f}({\bf u}(x,y,z,t), t)\rangle}
{|| \partial_x {\bf u}(x,y,z,t) ||^2}
\eeq
where
\beq
\langle {\bf u}(x,y,z,t)\cdot {\bf f}(x,y,z,t)\rangle = \int ({\bf u}\cdot {\bf f})(x,y,z,t)\, dx dy dz 
\eeq
is the usual Euclidean scalar product and $|| {\bf u}||^2=\langle {\bf u}\cdot {\bf u}\rangle$
the associated norm.
Since the velocity fields and the right hand side of the evolution equation vary in time,
 the phase speed calculated from (\ref{eq:c}) is usually also a function of time, $c(t)$.
This expression is the infinitesimal version of the shifts described in 
equation (2.13) of \citet{Mellibovsky2012} and has been discussed under various names by
various authors, see e.g. \citet{Smale1970,Rowley2003,Cvitanovic2012}.
We will use the term \emph{method of comoving frames}, since it reflects the underlying idea most directly:
we introduce a frame of reference that moves with the flow so that the advection is subtracted
and only the active part of the dynamics remains.

For travelling waves of the form ${\bf u}_{TW}(x-ct, y, z)$ the time evolution vector 
${\bf f}$ and the displacement vector $\partial_x {\bf u}$ are strictly parallel, so that the
advection speed determined by (\ref{eq:c}) agrees exactly with the phase speed of the wave,
i.e.\ ${\bf f}({\bf u}(x,y,z,t)) = c \partial_x {\bf u}(x,y,z,t)$.
We therefore checked our implementation of the method of comoving frames by computing the the advection
speed of a travelling wave (publicly available as TW2 in the database on \url{www.channelflow.org}).
and find that the result from the Newton algorithm agrees with our implementation using finite
differences to within a relative accuracy of $10^{-5}$.

\section{Flow states in the asymptotic suction boundary layer}
  The asymptotic suction boundary layer (ASBL) is the flow of an incompressible fluid over a flat plate into which
  the fluid is sucked with a constant homogeneous suction velocity $V_S$ \citep{Schlichting1987,Fransson2003}.
  We choose our coordinate system such that $x$ points downstream,  $y$ normal to the plate, and $z$ 
  in the spanwise direction. The free stream velocity scale is denoted $U_\infty$.
  Far away from any leading edge, the laminar flow profile is an exact solution to the incompressible Navier-Stokes equations,
  \beq {\bf u}(y) = (U_\infty ( 1-\mathrm e^{-y/\delta}), -V_S, 0), \eeq
  where $\delta = V_S / \nu$ is the displacement thickness.
  The laminar profile ${\bf u}(y)$ is translationally invariant in directions parallel to the wall.
  The Reynolds number is defined as $Re = U_\infty \delta / \nu = U_\infty / V_S$.
  Velocity, length and time are measured in units of $U_\infty$, $\delta$ and $t_0 = \delta/U_\infty$, respectively, with
  $t_0$ also referred to as advective time unit.
  
  We study the application of the method of comoving frames in the ASBL and in plane Poiseuille flow
  by direct numerical simulations, using
  channelflow \citep{channelflow}, as in our previous studies \citep{Kreilos2013,Zammert2013a}. Channelflow is 
  a pseudo-spectral code developed and maintained by John F. Gibson, 
  which expands the velocity field with Fourier-modes in the periodic directions and Chebyshev-Polynomials 
  in the wall-normal direction.
  In this section, we study the flow in the ASBL at Reynolds number $Re=500$
  in numerical domains of size 
  $L_x\times h\times L_z = 6\pi\times25\times3\pi$
  in units of $\delta$ with a resolution of $96\times193\times96$ modes,
  using periodic boundary conditions in the parallel to the walls and no-slip conditions at the walls.
  We have confirmed that a height of $25$ is sufficient by verifying that our results do not 
  change if a height of $50$ is used instead.
  The Reynolds number of $500$ in our calculations
  together with our chosen domain size is just large enough to make spontaneous decay
  from the turbulent state very unlikely and allows us to compute long turbulent trajectories.
  
\subsection{Downstream advection of the edge state}
  \label{sec:smalles}
\begin{figure}
    \centering
    \includegraphics{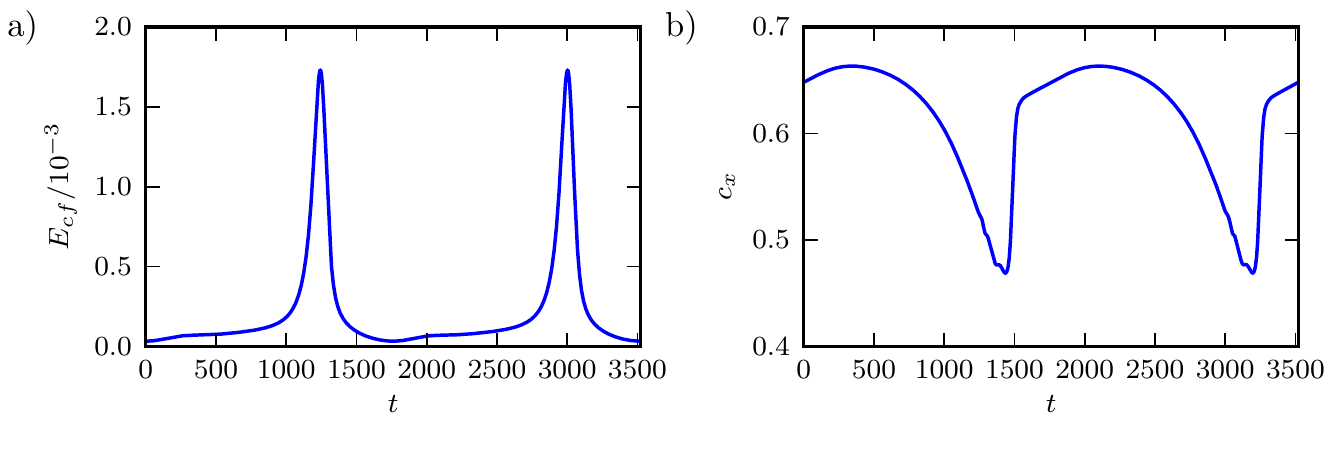}  
    \caption[]{Energy and advection velocity of the edge state in ASBL at $Re=500$.
      The left frame shows the clear increase in energy when the edge state goes through the 
      bursting phase. Studies of the flow field then show that afterwards it is displaced by half a box width, so that the full period are two bursting events in energy. The left frame shows the advection speed
      extracted by symmetry projection. It shows that as the energy goes up the structures slow down 
      noticeably.
      }
    \label{fig:edgestate}
  \end{figure}

  The edge state in the asymptotic suction boundary 
  layer is a periodic orbit with two different time scales \citep{Kreilos2013}.
  On a short time scale of about 15 advective time units,
  the dynamics is similar to that of a travelling wave. On longer time scales of about 1500
  advective time units one finds energetic bursts that are clearly visible in the 
  cross-flow energy%
  \footnote{The cross flow energy is a measure that has proven useful 
  in the characterization of the dynamics of shear flows. It contains only the energy in the 
  wall-parallel directions and is hence not dominated by the high energy content of the downstream 
components. It is defined by
$E_{cf} = \int_{V}\left({v^2 + w^2}\right) {\mathrm d}^3x/V$ where the integration is over
a box of volume $V$.}
  in figure~\ref{fig:edgestate}(a). The flow is composed of two counter-rotating downstream-oriented vortices and a 
  pair of low- and high-speed streaks.
  During the bursts, the streaks break up, and reappear shifted by half a box width with 
  respect to their original position.
  
  Observation of the dynamics during a burst is complicated by the advection of the structures 
  which happens on a much faster time scale than the bursts. Using (\ref{eq:c}) one can calculate
  an instantaneous advection speed $c_x$ in the downstream direction and go to a comoving frame of reference that removes the
  fast oscillations and reveals the slow dynamics underlying the bursts. 
  Figure  \ref{fig:edgestate}(b) shows the instantaneous advection velocity which is almost constant 
  during the low-energy phase of the state and drops noticeably as the energy increases
  and the state becomes more complicated.

\begin{figure}
   \centering
    \includegraphics{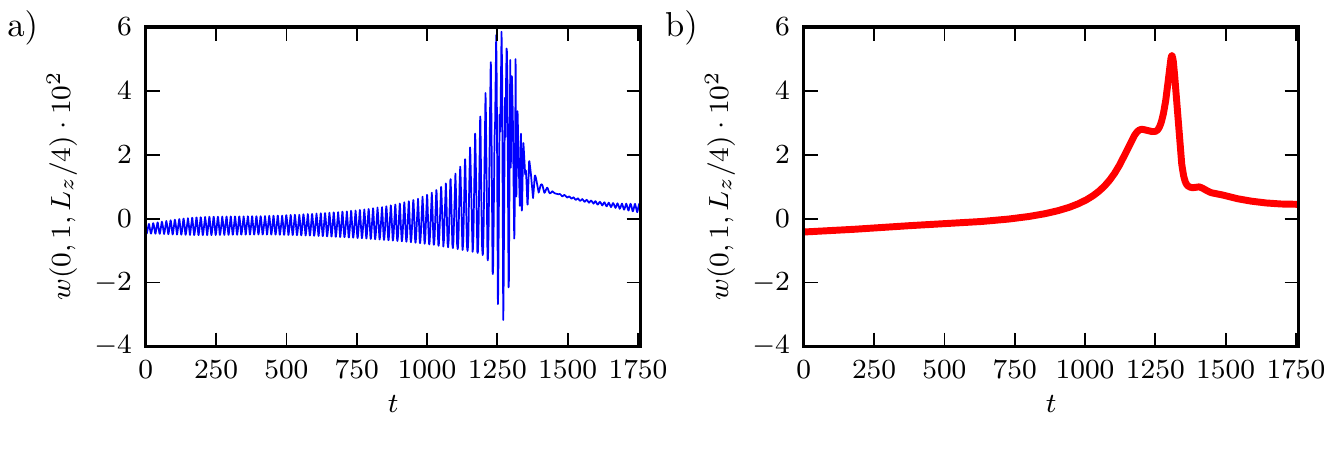} 
    \caption[]{\label{fig:edgestate_comparison}
      Comparison of the edge state dynamics in the laboratory frame and the comoving frame of reference. 
      Shown is the
      time evolution of a single spanwise velocity
      component, $w(0,1,L_z/4)$, as a function of time for one burst.
      In the laboratory frame of reference (left panel), the signal shows rapid oscillations, corresponding to the travelling-wave like behaviour of the edge state on short time
      scales, with the energetic burst visible as regions of larger amplitude oscillations.
      In the comoving frame of reference (right panel), the rapid oscillations are gone and the evolution of the component is rather smooth with an increase during the burst: the short-term 
      dynamics has been completely eliminated and only the long-term dynamics are visible.
      A movie of the time evolution that shows the differences much more clearly is provided as supplementary material.
    }
  \end{figure}

  The full power of the method of comoving frames and the different dynamics in the laboratory and the comoving frame of reference is
  most prominently visible in the movie provided with the online material. To give some indication of the differences, 
  we have extracted the spanwise velocity component at one point in the flow. The time trace in the labframe is shown in 
  figure~\ref{fig:edgestate_comparison}(a) and in the comoving frame of reference in figure~\ref{fig:edgestate_comparison}(b).
  The high-frequency jitter on the signal in the labframe is due to the rapid advection in the downstream direction.
  In the comoving frame this is removed entirely, so that the gradual built up of a strong spanwise velocity and its
  rapid break down stands out clearly. The movie then shows that the velocity field during the 
  low energy phase is dominated by down stream streaks with a weak sinusoidal modulation in the spanwise
  direction. As time goes on, the modulation amplitude increases, until the streaks break up. After the burst, the
  streaks and vortices form again, but displaced by half a box width in the spanwise direction.
  
  The calculations are in a domain with periodic boundary conditions in the downstream and the 
  spanwise directions. The flow naturally attains a discrete symmetry, a shift in the downstream direction
  by half a box length and a reflection on the midplane in the spanwise direction \citep{Kreilos2013}.
  With this discrete
  symmetry, the symmetry related advection velocities in the spanwise direction as calculated from
  (\ref{eq:c}) vanish exactly.
  For turbulent velocity fields this symmetry is broken and spanwise
  components do appear, as we discuss in \S \ref{sec:turb}.
  
\subsection{Spanwise advection of localized edge states}

  The discrete symmetry involving a spanwise reflection is broken if the domain is extended wide enough 
  in the spanwise direction so that
  the state localizes. Edge states in this setting have been calculated in \citep{Khapko2013,Khapko2014}; 
  their dynamics is similar to the one in periodic domains, and shows long calm phases that are interrupted by violent bursts, during which the structures break up and reform at a position that is shifted in the spanwise direction.
These studies also show that different patterns in the spanwise displacement are possible: states may always be shifted to the right or the left, they may alternate between left and right
  shifts or they may even follow irregular patterns of left and right shifts. We here focus on a state that always
  shifts to the left. 
   
  \begin{figure}
    \centering
    \includegraphics{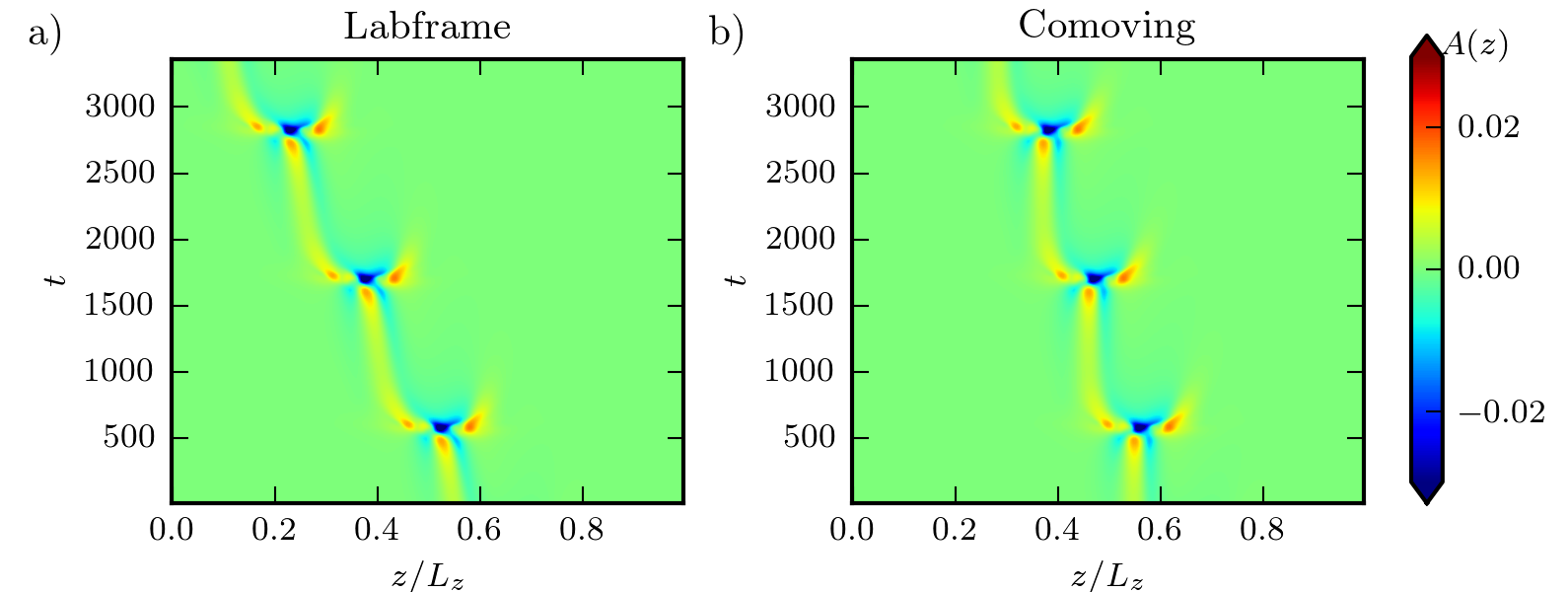}
    \caption{\label{fig:Az_es}
      Space-time plot of the $x$-averaged wall-normal velocity $A(z,t)$, eq (\ref{eq:A}), for the
      left jumping edge state in a wide domain \citep{Khapko2013}.
      The edge state is localized and during the calm phases between two bursts it consists of a pair
      of downstream-oriented counter-rotating vortices, visible by the alternating pattern of fluid
      moving up (red) and down (blue). At the bursts the structures break up and reform at a different
      spanwise location.
      a) In the laboratory frame, in the calm phases the red and blue lines are tilted,
      reflecting a spanwise drift.
      b) In the comoving frame of reference, the red and blue regions are aligned along the time-axis during
      the calm phases, indicating that the drift has been removed. However, the spanwise displacement at the bursts
      remains, indicating that it is not caused by advection but by internal dynamics of the flow.
    }
  \end{figure}
  
  In order to obtain a 2-d representation of the four-dimensional space-time evolution of the flow 
  state we pick the wall-normal velocity at the height $\delta$ as a good indicator of up- and downwelling motion
  that can be connected to vortices. Since the domains are relatively short, we average this quantity in the
  downstream direction. The resulting observable
   \beq
  A(z,t) = \langle v(x,y=\delta,z,t) \rangle_x
  \label{eq:A}
  \eeq 
  depends on the spanwise coordinate $z$ and time $t$ and can be represented in 2-d color plots. For
  three periods of the left-shifting state we find the representation shown in figure~\ref{fig:Az_es}.
  In the laboratory frame of reference (left panel), we see
  that there is an alternating pattern of fluid moving up and down (red and blue), corresponding to alternating 
  streamwise vortices.
  At the bursts, the structures break up and are reformed almost symmetrically left and right, of which only the
  left one survives.
  In between the bursts the structures slowly and constantly drift to the left, as indicated by the finite slope of the
  lines.
  In the right panel, in the comoving frame
  of reference, the drift is completely removed and the structures are stationary in the calm phase, while they 
  are still displaced at the bursts.
  This clearly shows that the method of comoving frames is able to separate the advective part of the time evolution
  (i.e.\ the drift) from the dynamically active and relevant part, in this case the break-up and displacement of the
  structures.

\subsection{Spanwise drift of turbulent states}
  \label{sec:turb}
\begin{figure}
    \centering
    \includegraphics{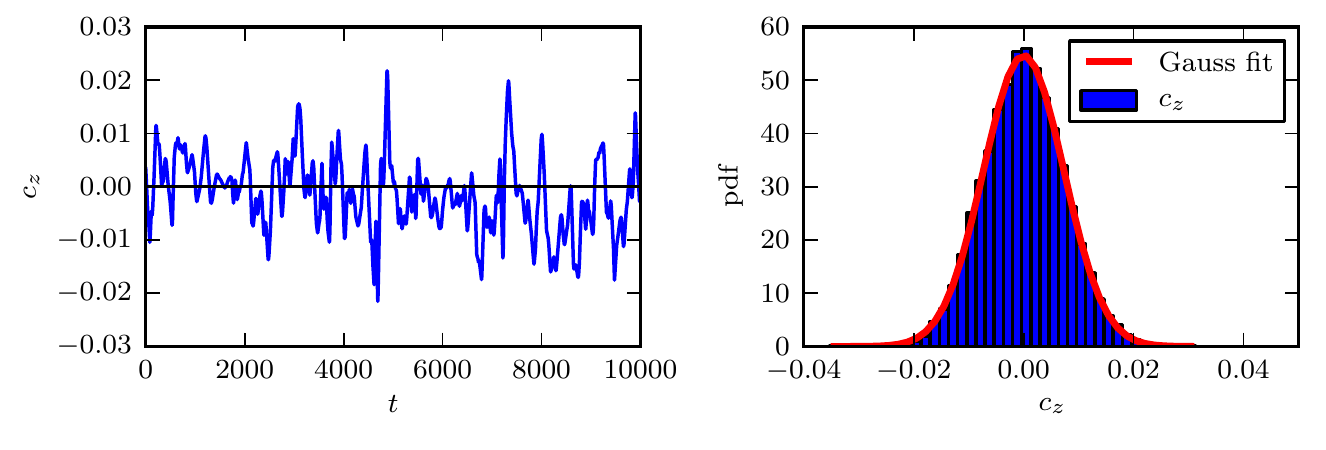} 
    \caption{Transverse advection velocity for a turbulent trajectory in the ASBL.
      a) The spanwise advection velocities $c_z$ are small and fluctuate randomly.
      b) The distribution of the advection velocities is almost perfectly Gaussian, 
      indicating that it mainly consists of random noise.
    }
    \label{fig:example_vz}
  \end{figure}

  We now turn to turbulent flows and an analysis of spanwise shifts within the same computational domain
  as in \S\ref{sec:smalles}. 
  The general expression (\ref{eq:c})
  for the phase speed with $\partial_x{\bf u}$ replaced by $\partial_z{\bf u}$ will not vanish in general
  (unless there are discrete symmetries as mentioned before). However, one would expect this contribution to 
  be small and more or less uncorrelated. Indeed, the time traces in figure~\ref{fig:example_vz}(a) 
  show a  strongly fluctuating signal with a small amplitude and a probability density function (pdf) that is well
  approximated by a Gaussian (figure~\ref{fig:example_vz}(b)).
  
\begin{figure}
    \centering
    \includegraphics{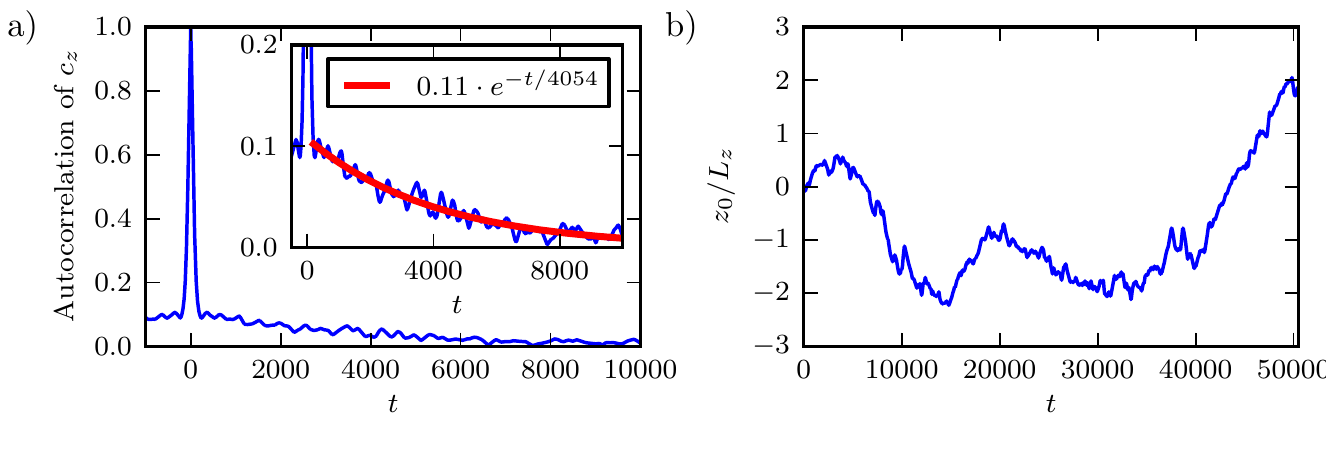} 
    \caption[]{Correlation function and integrated displacements in spanwise direction in the ASBL.
    The correlation function of the spanwise advection speed (left frame) shows a strong central peak, 
    indicative of weak correlations, and a broad background reaching out to very long times. 
    The inset shows that this
    tail is well captured by an exponential distribution with a characteristic time of more than
    4000 natural units. 
    The frame on the right shows the integrated advection speed, or displacement, in the spanwise direction. 
    The long correlations show up as modulations over very long periods. Note that the time scale covered 
    is five times that shown in figure~\ref{fig:example_vz}(a).}
    \label{fig:autocorrelation}
  \end{figure}

  The correlation function of the advection speeds is shown in figure~\ref{fig:autocorrelation}(a). 
  It is based on data from 14 turbulent trajectories, each integrated for 200 000 time units.
  Initially, the correlation function falls of rather quickly, but then develops a wide background. 
  As the inset shows, this background can be well approximated by an exponential form with a
  characteristic time of more than 4000 advective time units. Armed with the information that this 
  correlation time is so long, one can begin to see evidence for it in figure~\ref{fig:example_vz}(a), where the average 
  of the fluctuations over the first 3000 time units is slightly above 0, whereas it is 
  below zero for the remaining 7000 time units. The consequences of this are that the integrated 
  advection speed, i.e. the distance over which the velocity field are displaced,
\begin{equation}  
z_0(t) = \int_0^t c_z(\tau) d \tau ,
  \label{eq:z0}
\end{equation}
  first increases  and then decreases. This is shown in figure~\ref{fig:autocorrelation}(b).

\begin{figure}
   \centering
   \includegraphics{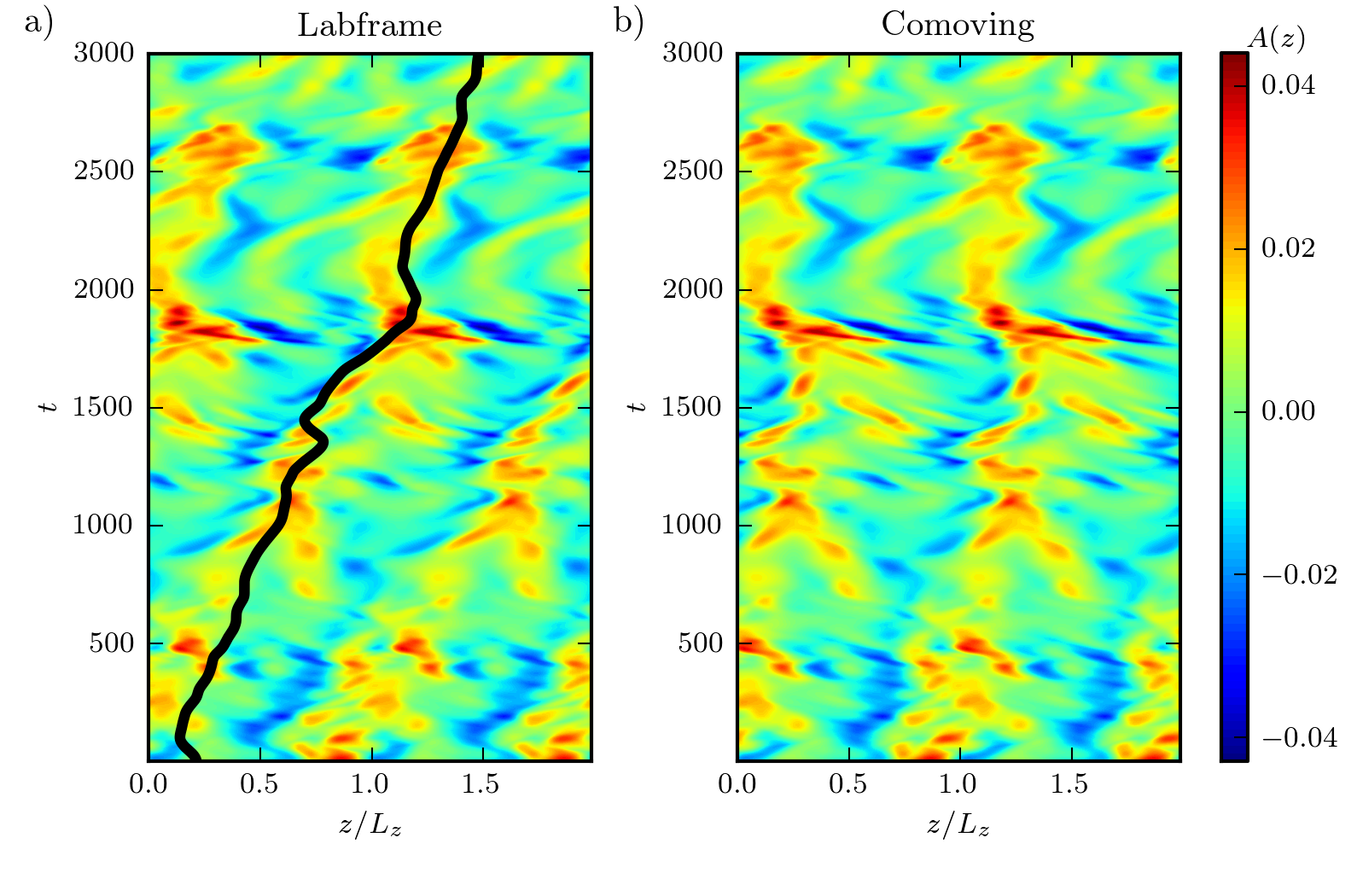}
   \caption{\label{fig:Az}
     Space-time plot of the $x$-averaged wall-normal velocity $A(z,t)$, eq (\ref{eq:A}), demonstrating the effect of
     the method of comoving frames on a turbulent trajectory in the ASBL.
     The width of the box is twice the computation domain in order to highlight advection across the spanwise
     boundaries.
     For most times, there are two regions, predominantly blue and red, respectively, corresponding 
     to one pair of large-scale downstream vortices.
     a) In the laboratory frame, the blue and red regions are tilted, reflecting a gradual spanwise drift. The thick black line is
     the integrated advection velocity, and it falls along a prominent ridge in the structures. 
     b) In the comoving frame of reference, where the drift is subtracted, the two colored regions are 
     essentially stationary and aligned along the time-axis.
   }
  \end{figure}

The effect of the transverse drift and its removal on the flow patterns is shown in figure~\ref{fig:Az},
in the reduced representation (\ref{eq:A})  of the velocity fields. 
  Regions with strongly positive or negative components in the normal direction
  show up as ridges that meander in space and time. This is particularly noticeable in the labframe (left frame).
  The black line on top of the figure is the integrated spanwise drift (\ref{eq:z0}): it is perfectly aligned with the ridges in $A(z,t)$ and underlines the drift to the right. 
  In the comoving frame of reference (right panel), the structures are then shifted by this amount in the spanwise
  direction, and the structures are more parallel to the vertical axis. 
   
\begin{figure}
   \centering
   \includegraphics{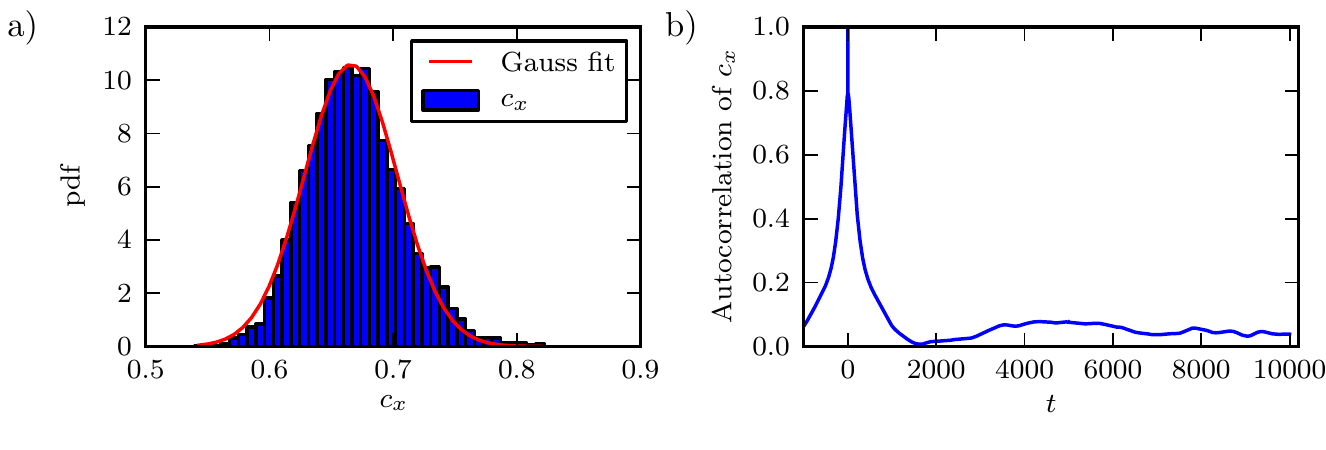}
   \caption{Pdf and autocorrelation function for the downstream advection velocity in the ASBL. In contrast to the spanwise
    advection, figure~\ref{fig:autocorrelation}, it drops rather quickly to zero and does not show the long tails 
    noted above.}
    \label{fig:autocorrelationx}
  \end{figure}

  In view of the long time correlations found in the spanwise direction, we also checked the 
  downstream advection velocity.
  The downstream component fluctuates around a mean advection velocity 
  $\overline{c}_x\approx 0.667$. 
  The pdf is close to Gaussian with a weak but noticeable asymmetry to larger speeds,
  see igure~\ref{fig:autocorrelationx}(a).   
  The autocorrelation function of the fluctuations around the mean, 
  figure~\ref{fig:autocorrelationx}(b),  falls off
  rapidly, but is significantly wider than the one for the spanwise shifts. The background is broader
  but less well characterized by a single exponential than in the case of the spanwise shift. 
  
\section{Turbulent flow states in plane Poiseuille flow}
In this section we apply the method of comoving frames to plane Poiseuille flow. As shown in 
\citep{Toh2003,Zammert2013a}, plane Poiseuille flow has exact coherent structures that are remarkably 
similar to the ones found in the ASBL, and their phase speeds in the streamwise and spanwise
directions can be determined as in the case of the ASBL. We here focus on a turbulent state and
its transverse drift.

We perform direct numerical simulations of plane Poiseuille flow with constant bulk velocity at $Re=3000$, based on the laminar center-line velocity $U_0$ 
and the channel half-width. The laminar flow profile is given by ${\bf u}(y) = (1-y^{2}, 0, 0)$, 
so that the bulk velocity becomes $U_{bulk} = (2/3) U_0$. For the computations we use a 
domain of size $L_{x} \times L_{y} \times L_{z} = \pi  \times 2 \times 0.4\pi$,  a resolution of $32 \times 65 \times 32$ modes, and Gibson's code channelflow \citep{channelflow}.

\begin{figure}
   \centering
   \includegraphics{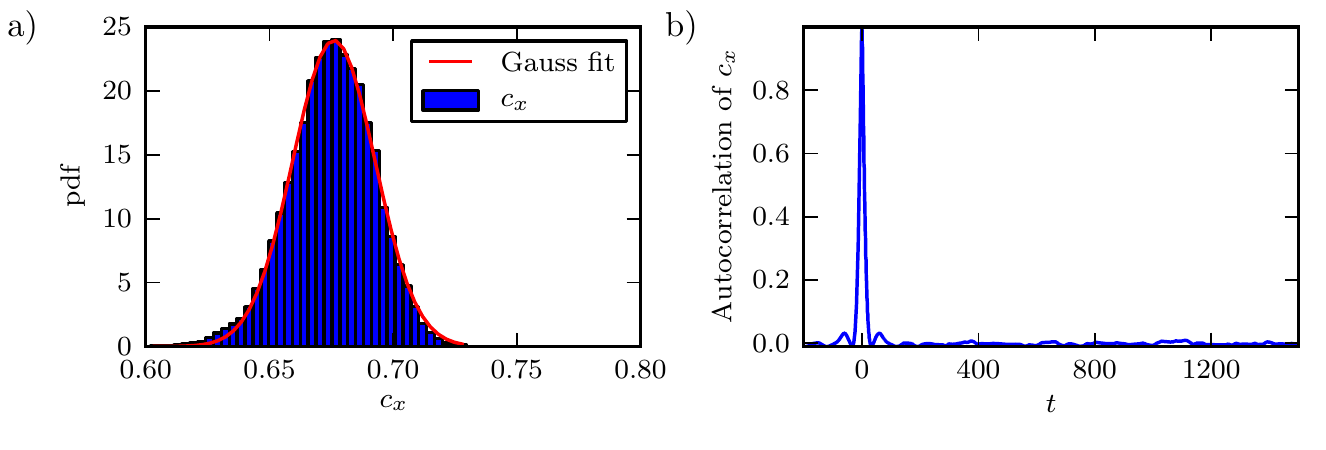}
   \includegraphics{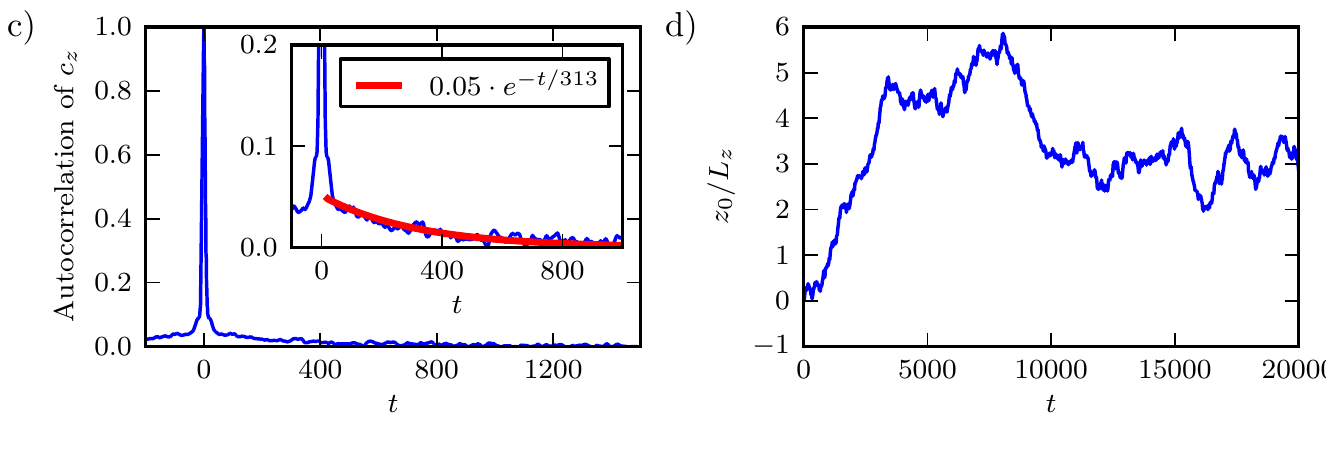} 
   \caption{Application of the method of comoving frames to plane Poiseuille flow at $Re=3000$.
    (a) The streamwise advection velocity fluctuates around a mean value of $0.675$ what is close to the bulk velocity. The distribution of the velocities is almost gaussian.
    (b) The correlation function of the downstream advection velocity quickly drops to zero.  
    (c) The correlation function of the spanwise advection speed shows a strong central peak, indicative of weak
   correlations, and a broad background reaching out to long times. The inset shows that this
    tail is well captured by an exponential distribution with a characteristic time of about 300 time units. 
    In (d) the the integrated advection speed, or displacement, in the spanwise direction is shown. }
    \label{fig:ResultsPPF}
  \end{figure}

As in \S\ref{sec:turb} we study the spanwise and streamwise drifts. 
To calculate the correlation function of the advection speeds data from 20 turbulent trajectories, each integrated for 50 000 time units $d/U_{0}$
is used. The average of the downstream advection velocity $c_x$ is $0.675$ and therefore slightly larger than the bulk velocity. As in the ASBL, the probability density function
of $c_x$ in figure \ref{fig:ResultsPPF}(a) is well approximated by a Gaussian and the correlation function
of the downstream advection velocities, shown in figure \ref{fig:ResultsPPF}(b), quickly drops to zero.

The spanwise advection velocities are also Gaussian distributed with a mean value of zero (not shown).
The correlation function for the spanwise velocities in figure \ref{fig:ResultsPPF}(c) also shows a 
wide background as in the case of the ASBL. An exponential fit yields a
characteristic time of about 300 time units. This weak long-ranged correlation can be seen in the 
integrated advection speeds $z_{0}$ (\ref{eq:z0}) in figure~\ref{fig:ResultsPPF}(d).
The plot reveals several segments in which the flow drifts in the same direction. They are a few hundred time units 
long, consistent with the correlation time of about 300. 

The different transverse correlation times are unusual. 
Compared to the ASBL the characteristic times of the background correlation differ by a factor of 13. Yet the 
coherent structures studied by \citet{Toh2003} and \citet{Zammert2013a} show bursts on time scales that are similar.
The difference presumably comes about because the coherent structures in plane Poiseuille flow are localized near one wall, whereas the determination
of the advection speed involves an integration over the entire domain. 
Uncorrelated motions close to the upper and lower walls will thus reduce the correlation time.

\section{Concluding remarks}
The examples show that one of the appealing features of the transformation (\ref{eq:c}) is the local
and instantaneous removal of a downstream advective component in the time evolution. The speed
by which this component is removed depends on the state and hence varies in time.
However, once this component is removed, the changes in vortices and streaks stand out much more
clearly and become accessible to further analysis. 

It is possible to extend this analysis from local in time also to local in space, and to find advection speeds
that focus on particular features of the flow. The key to this is the scalar product that
enters the projection to obtain  eq.~(\ref{eq:c}). In the calculations shown here we
use the usual Euclidian scalar product with the scalar product of all components integrated over all of space,
eq.~(2.6).
For other situations and applications, suitable adaptations of the method are possible. For instance, 
if not all velocity components are available, one could base the projection on a subset of components. 
Also, if the velocity fields are only known in a subset of the space, the integration may be limited to that
subset. Extensions in this direction are currently being explored.

Another extension of the method uses different representations of the velocity field, e.g.
expansions in spectral modes and restrictions in the scalar product to subsets of the modes.
For instance, if the velocity fields are given in a Fourier representation, one can extract phase speeds
for individual Fourier components. \citet{DelAlamo2009} used such an idea in their intriguing 
discussion of the effects of Taylors frozen flow hypothesis \citep{Taylor1938,Townsend1980,Zaman1981}
on turbulent spectra: they defined a wavenumber dependent phase speed (their equation (2.4)) that is equivalent to 
(\ref{eq:c}) when applied to a single spanwise Fourier mode.

The method for the extraction of symmetry related motions and the expression for the advection
speed given in equation (\ref{eq:c}) are easy to calculate and to extract from both numerical and 
experimental data. For the use with experimental data, the time-evolution vector is replaced by finite
differences between velocity fields at different times, and the spatial derivative can be approximated
by finite differences in space. With a sufficiently good spatial and temporal resolution, the results
are indistinguishable from full analytical approximations. 
If the time steps are longer it may be advisable to turn to the optimization methods described in 
\citet{Mellibovsky2012}, where the velocity field ${\bf u}(x,y,z,t)$ is used as a template and a suitable
shift is determined by minimizing 
$|| ({\bf u}(x,y,z,t+\Delta T)-{\bf u}(x-\Delta x,y,z,t)||^2$. The shift and advection speed determined
by this method and the one from eq. (\ref{eq:c}) become equivalent in the limit of small $\Delta t$
where Taylor expansions are possible.

Continuous symmetries appear in many fields, and various methods for separating shifts along the symmetry
axis from dynamical changes have been developed (see \citet{chaosbook,Froehlich2011,Cvitanovic2012}). 
In the fluid mechanical context they have
appeared in connection with derivations of low-dimensional models using proper orthogonal
decomposition \citep{Rowley2000,Rowley2003}. Their relevance for the detection of relative
periodic orbits, including a successul application of the method of slices, which allows an exact
symmetry reduction, has been emphasized in \citet{Willis2013}. 

The method of comoving frames described in \S2 helps in removing some of the advective and dynamically less relevant
components, but it does not solve all problems related to finding relative periodic orbits, since it is not
a general symmetry reduction scheme.
For instance, the state shown in the first example in \S3 passes
through an intermediate state where the flow fields are shifted by half a box width in the spanwise 
direction. This shift is a dynamic one, not related to advection, and hence not removed. 
Therefore, for a periodic orbit calculation, one would
still have to take the initial state and its images under discrete symmetries and check whether the 
flow returns to the symmetry related copy. The situation gets even more complicated for the  states in the wider 
domains as discussed in \citet{Khapko2013,Khapko2014}:
For instance, the states that steadily
move to the left do repeat after a suitable shift in the spanwise direction,
but as shown in \S3.2 this translation is again not removed by the continuous shifts since it is a result of the
changes in the flow fields.

Nevertheless, judging by the examples we have analyzed so far, the
removal in particular of the downstream advection reveals underlying coherent structures,
or candidates for coherent structures, more clearly than the laboratory frame dynamics.

\subsection*{Acknowledgements}
  We thank John F Gibson for providing and maintaining the open source Channelflow.org code,
  Marc Avila for discussions that motivated this investigation, Predrag Cvitanovi\'c for extensive
  exchanges on continuous symmetries, Hannes Brauckmann and Matthew Salewski for 
  comments and Francesco Fedele for encouragement. 
  This work has been supported in part by the Deutsche Forschungsgemeinschaft 
  within Forschergruppe FOR1182.
  
\bibliographystyle{jfm}
\bibliography{library}

\end{document}